\documentclass[iop]{emulateapj}
\usepackage{graphicx}
\usepackage{amssymb}
\usepackage{epstopdf}
\usepackage{amsmath}
\usepackage{mathrsfs}
\usepackage{natbib}
\usepackage{enumitem}
\usepackage{setspace}
\usepackage{float}
\usepackage{times}
\usepackage{subfigure}
\defcitealias{Einasto2012a}{E12a}
\defcitealias{Einasto2012b}{E12b}
\defcitealias{Cohen2014}{C14}

\slugcomment{Accepted for publication in The Astrophysical Journal}

\begin{document}

\shorttitle{Star Formation and Superclusters}
\shortauthors{Cohen et al.}

\title{Star Formation and Supercluster Environment of 107 Nearby Galaxy Clusters}

\author{Seth A. Cohen, Ryan C. Hickox, Gary A. Wegner}
\affil{Department of Physics and Astronomy, Dartmouth College, 6127 Wilder Laboratory, Hanover, NH 03755, USA}
\author{Maret Einasto, Jaan Vennik}
\affil{Tartu Observatory, 61602 T\~{o}ravere, Estonia}

\begin{abstract}
We analyze the relationship between star formation (SF), substructure, and supercluster environment in a sample of 107 nearby galaxy clusters using data from the Sloan Digital Sky Survey. Previous works have investigated the relationships between SF and cluster substructure, and cluster substructure and supercluster environment, but definitive conclusions relating all three of these variables has remained elusive. We find an inverse relationship between cluster SF fraction ($f_{SF}$) and supercluster environment density, calculated using the galaxy luminosity density field at a smoothing length of 8 $h^{-1}$ Mpc (D8). The slope of $f_{SF}$ vs. D8 is $-0.008 \pm 0.002$. The $f_{SF}$ of clusters located in low-density large-scale environments, $0.244 \pm 0.011$, is higher than for clusters located in high-density supercluster cores, $0.202 \pm 0.014$. We also divide superclusters, according to their morphology, into filament- and spider-type systems. The inverse relationship between cluster $f_{SF}$ and large-scale density is dominated by filament- rather than spider-type superclusters. In high-density cores of superclusters, we find a higher $f_{SF}$ in spider-type superclusters, $0.229 \pm 0.016$, than in filament-type superclusters, $0.166 \pm 0.019$. Using principal component analysis, we confirm these results and the direct correlation between cluster substructure and SF. These results indicate that cluster SF is affected by both the dynamical age of the cluster (younger systems exhibit higher amounts of SF); the large-scale density of the supercluster environment (high-density core regions exhibit lower amounts of SF); and supercluster morphology (spider-type superclusters exhibit higher amounts of SF at high densities).
\end{abstract}


\section{Introduction}
\label{sec:IntroSec}

The effects of galaxy cluster mergers on star formation (SF) have begun to be better understood in recent years, adding depth to the relationships found in relaxed clusters between SF and clustercentric distance and local density (e.g., \citealt{Dressler1980}; \citealt{Cohen2014}, hereafter C14; \citealt{Cohen2015}; and many others). While some studies find no relationship between cluster merger activity and SF in specific clusters \citep[e.g.,][]{Metevier2000, Ferrari2005, Braglia2009, Hwang2009, Kleiner2014}, many others report such a relationship \citep[e.g.,][]{Knebe2000, Cortese2004, Ferrari2005, Johnston2008, Bravo2009, Braglia2009, Hwang2009, Ma2010, Wegner2011, Wegner2015, Sobral2015, Girardi2015, Stroe2015}. Indeed, \citetalias{Cohen2014} and \citet{Cohen2015} found that SF is statistically correlated to cluster substructure in studies of large numbers of clusters: in general, clusters with more substructure exhibit greater levels of SF.

Recent studies have investigated the relationship between cluster substructure and supercluster environment \citep[][hereafter E12b]{Einasto2015, Krause2013, Einasto2012b}. In particular, \citetalias{Einasto2012b} found that clusters in superclusters are more likely to have substructure than those that are isolated, though the correlation discussed in the paper is weak. Studies have also begun to probe the connection between supercluster environment and SF \citep[e.g.][]{Costa2013, Luparello2013, Lietzen2012}. In voids, there is a general consensus that this lower-density large-scale environment only weakly affects galaxy properties, which depend more strongly on local environment \citep{Grogin2000, Rojas2005, Patiri2006, Wegner2008, Hoyle2012, Kreckel2011, Kreckel2012}.

In superclusters, \citet{Einasto2014} recently showed that supercluster morphology is important in shaping the properties of galaxies: higher levels of SF are found in galaxies in spider-type superclusters than filament-type superclusters. Simulations by \citet{Aragon2014} suggest that the quenching of SF in clusters depends on the geometry of the large-scale surrounding structure. This supports observational work by \citet{Einasto2014}: spider-type superclusters have richer inner structure and larger numbers of filaments connecting galaxy clusters than do filament-type superclusters.

However, a definitive relationship between supercluster environment and SF has yet to be shown. We seek to develop a coherent picture connecting these four variables: cluster star-forming fraction ($f_{SF}$), amount of cluster substructure, supercluster environment density, and supercluster morphology. This paper considers the correlations between these parameters, focusing in particular on the pairwise comparison between supercluster environment and SF. Furthermore, we seek to confirm the pairwise comparisons involving cluster substructure and cluster SF, and cluster substructure and supercluster density. Finally, we investigate a potential multi-dimensional correlation among the three non-morphological variables. In \S\ref{sec:DataSec}, we introduce our cluster sample and discuss methods for determining substructure, SF, and supercluster properties; \S\ref{sec:ResultsSec} enumerates our results; and the implications of our findings we discuss in \S\ref{sec:DiscussionSec}. Throughout our analysis we assume a standard cosmology of $H_{0} = 100\:h\:\textnormal{km}\:\textnormal{s}^{-1}\:\textnormal{Mpc}^{-1}$, $\Omega_{\textnormal{m}} = 0.27$, and $\Omega_{\Lambda} = 0.73$.

\section{Data and Methods}
\label{sec:DataSec}

In this section, we describe our cluster and supercluster samples. We also explain our methods for calculating SF and substructure properties of clusters. Finally, we introduce our use of principal component analysis (PCA) in determining relationships between SF, substructure, and large-scale environment.

\subsection{Cluster sample}
\label{sec:SampleDataSec}

We use the sample of rich clusters from the group catalogue of \citet{Tempel2012}, which is based on the SDSS DR8 spectroscopic data \citep{Aihara2011}. Using SDSS data, \citet{Tempel2012} identified 77,858 groups and clusters using the friends-of-friends (FoF) method \citep{Zeldovich1982, Huchra1982}. \citealt{Einasto2012a} (hereafter E12a) used the subsample of rich clusters with at least 50 members in the redshift interval $0.04 \leq z \leq 0.1$ to determine the substructure properties of the clusters. They found that 90 of these clusters contain substructure and 17 do not. In the present paper, we use this cluster sample, previously analyzed by \citetalias{Cohen2014}. We obtain galaxy stellar masses from the Max Planck Institute (MPA)/Johns Hopkins University (JHU) VAGC \citep{Tremonti2004}, and both observed and estimated total \emph{r}-band luminosities ($L_{obs}$ and $L_{tot}$, respectively) from the catalogue of \citet{Tempel2012}.

As the SDSS data are flux-limited, the FoF method potentially suffers from the bias of fainter galaxies vanishing as distance increases. This leads to differences in the luminosities of member galaxies between nearby and more distant groups. \citet{Tempel2012} partly corrected for this effect by determining a relationship between distance and the linking length used in their FoF algorithm, and then applying this relation when selecting groups at different distances. They note that, by applying this correction, their final group catalogue is quite homogeneous in richness, size, and velocity dispersion, regardless of distance. However, we note that this does not correct for the fact that groups of a given richness at lower redshift are less luminous than those of the same richness at higher redshift. This is because, spectroscopically, fainter galaxies are more easily detected in the SDSS -- and thus included as group members -- at lower redshift than higher redshift. Despite this bias, in \S\ref{sec:PairwiseSec}, we discuss how our results are not affected by differences in cluster luminosity.

To further alleviate the biases inherent in flux-limited surveys, following the prescription in \citetalias{Cohen2014}, we study only those galaxies with $M^{0.1}_{r} < -20.5$. The determination of this absolute magnitude cut follows the methods of \citet{Hwang2009}. A galaxy's \emph{r}-band absolute magnitude is calculated from its apparent magnitude $m_{r}$ via 
\begin{equation}
\label{equ:MrEqu}
M^{0.1}_{r} = m_{r} - DM - K(z) - E(z),
\end{equation}
where $m_{r}$ is corrected for extinction; $DM \equiv 5\:\textnormal{log}(D_{L}/10\:\textnormal{pc})$ and $D_{L}$ is a luminosity distance; $K(z)$ is a \emph{K}-correction \citep{Blanton2007} to a redshift of 0.1, denoted by the superscript; and $E(z)$ is an evolution correction defined by $E(z) = 1.6(z - 0.1)$ \citep{Tegmark2004}.  Extinction-corrected magnitudes and \emph{K}-corrections are collected from the NYU Value-Added Galaxy Catalogue \citep[VAGC;][]{Blanton2005, Padmanabhan2008}.

\subsection{Star formation and substructure determinations}
\label{sec:SFStrucDataSec}

\citetalias{Cohen2014} determined which galaxies are star-forming using the detection of H$\alpha$ emission, defined as the measurement of an equivalent width of at least 3 \AA\:\citep[a compromise between, e.g.,][]{Ma2008, Balogh2004, Rines2005}.  Relevant equivalent width and flux measurements were retrieved from the MPA/JHU VAGC \citep{Tremonti2004}. When possible, they also used the BPT diagram \citep{Baldwin1981} that uses the emission line ratios $\log([\textnormal{OIII}]\lambda5007/\textnormal{H}\beta)$ vs. $\log([\textnormal{NII}]\lambda6583/\textnormal{H}\alpha)$ to separate star-forming galaxies from AGN and LINERs \citep{Kauffmann2003, Kewley2001}; the latter two types we remove from our analysis. A cluster's $f_{SF}$ is defined as the number of star-forming galaxies divided by the total number of galaxies in the cluster.

Cluster substructure properties were determined by \citetalias{Einasto2012a} using multidimensional normal mixture modelling via the \emph{Mclust} package for classification and clustering \citep{Fraley2006}. \emph{Mclust} assigns each member galaxy to a component, thus determining the number of components in each cluster. \citetalias{Einasto2012a} also analyzed the substructure properties of our clusters using the Dressler-Shectman (DS or $\Delta$) test \citep{Dressler1988}. In short, for each cluster, this test measures how each galaxy's local kinematics differ from the kinematics of the cluster as a whole. The results of the test are then calibrated using Monte Carlo simulations to determine a \emph{p}-value, the probability that any observed substructure is due to chance. Thus, smaller \emph{p}-values indicate higher probabilities of substructure. Please see \S3.2 in \citetalias{Einasto2012a} for more details on the $\Delta$ test and its calibration.

\subsection{Large-scale environment of clusters}
\label{sec:SuperclustDataSec}

Most clusters belong to a supercluster, and these superclusters are characterized by their total luminosity, richness, and morphology \citepalias{Einasto2012b}. To demarcate superclusters, we use the methods of \citetalias{Einasto2012b}, who calculated the galaxy luminosity density field and determined the luminosity distribution of galaxies. Supercluster membership was determined at the smoothing length of 8 $h^{-1}$ Mpc (hereafter D8), and the density $\textnormal{D}8 = 5$ (in units of mean density, $\ell_{\mathrm{mean}} = 1.65 \cdot 10^{-2} \frac{10^{10} h^{-2} L_{\odot}}{(h^{-1} \textnormal{Mpc})^3}$) is used to separate supercluster environments from the field \citep{Liivamagi2012}. Furthermore, as determined in \citet{Einasto2007}, $\textnormal{D}8 \approx 8$ separates the high-density cores of superclusters from their outskirts. We direct the reader to Appendix B in \citetalias{Einasto2012b} and references therein for more details on these density calculations. We note that a correlation exists between D8 and redshift for our sample clusters. However, any evolution in redshift should be minimal within the redshift range we study, and this bias should not affect our conclusions.

\begin{figure}
\begin{center}
\subfigure{\includegraphics[scale=0.63]{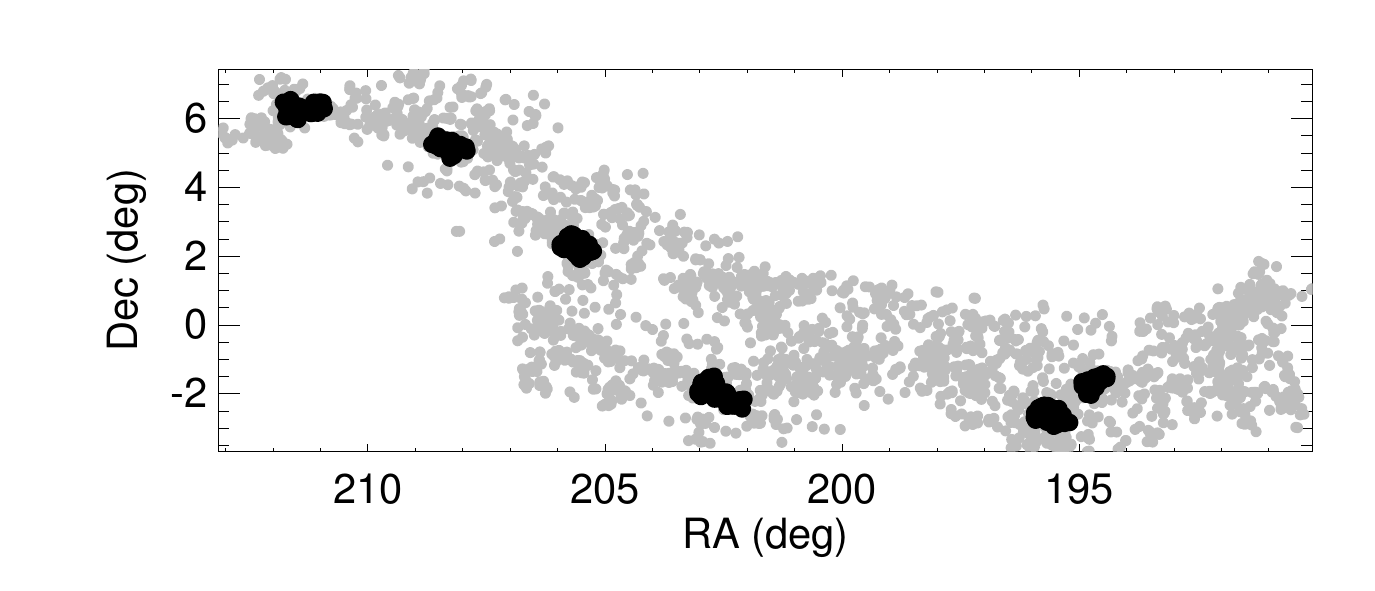}}\\
\subfigure{\includegraphics[scale=0.63]{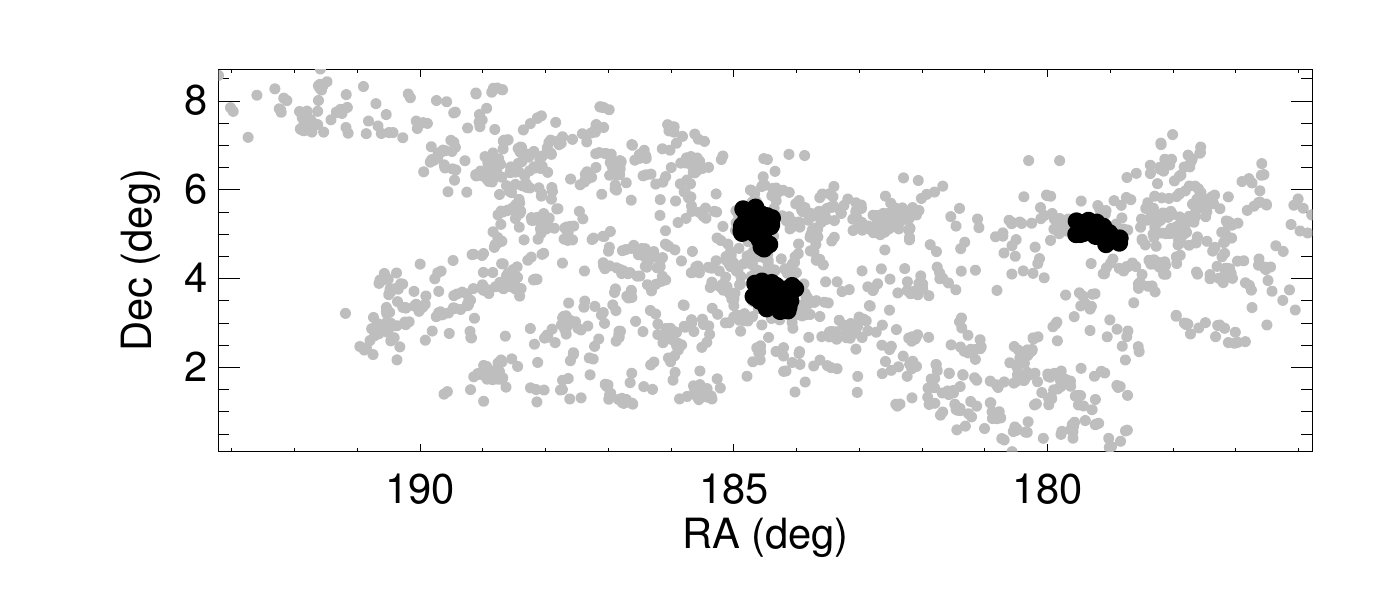}}
\caption{Examples of a filament supercluster (top) and a spider supercluster (bottom). These are the richest superclusters of the Sloan Great Wall, SCl~027 and SCl~019, respectively (see \citealt{Einasto2014} for details). Black circles denote galaxies in clusters of at least 50 members, and gray circles represent other galaxies in the supercluster.}
\label{fig:SClsFig}
\end{center}
\end{figure}

Supercluster morphology is determined by the four Minkowski functionals \citepalias{Einasto2012b}, which are proportional to an enclosed volume, the area of the surface surrounding it, the integrated mean curvature of this surface, and its integrated Gaussian curvature. The first three functionals describe the overall structure of a supercluster via two shapefinders (planarity and filamentarity) and their ratio (shape parameter). The fourth functional describes a supercluster's inner structure. This methodology divides superclusters into four morphologies, based on the Minkowski functions and visual appearance: spiders, multispiders, filaments, and multi-branching filaments \citep{Einasto2011}. For simplicity, in this work, we combine these classifications into two main types: spiders, which exhibit one or more high-density clumps of clusters connected by many galaxy chains; and filaments, in which high-density clumps or cores are connected by a small number of galaxy chains. Please see Appendix C in \citetalias{Einasto2012b} and references therein for details on these morphology calculations. Figure~\ref{fig:SClsFig} shows an example of a filament supercluster (top) and a spider supercluster (bottom). Galaxies in clusters of at least 50 members are shown in black, and other galaxies in the supercluster are shown in gray.

\subsection{Principal component analysis methods}
\label{sec:PCADataSec}

PCA has been widely used in astronomy for a number of purposes (see \citealt{Einasto2011} for references). PCA transforms variables of interest to a new coordinate system whose new variables are known as the principal components (PCs) of the data. These PCs are linear combinations of the original parameters, and they illustrate the variable(s) along which the original data has the most variance. The original data varies most when projected along the first PC; the direction of the second PC indicates the direction of the second greatest variance; etc. We normalize and centralize our parameters by dividing each by its standard deviation and centering each on its mean. We use PCA to investigate how several variables are potentially correlated with cluster SF. In particular, we focus not only on the two variables discussed in this work -- cluster substructure and supercluster environment -- but also include total cluster halo mass via two proxies, cluster \emph{r}-band luminosity (both $L_{obs}$ and $L_{tot}$) and total cluster stellar mass \citep[$M_{*}$; e.g.,][]{Yang2007, Andreon2010, Gonzalez2013}.

\section{Results}
\label{sec:ResultsSec}
\subsection{Pairwise comparisons}
\label{sec:PairwiseSec}

\begin{deluxetable}{cccccc}
\tablecolumns{6}
\tablewidth{0pc}
\tablehead{
\colhead{} & \colhead{Supercluster} & \colhead{Environmental} & \colhead{} & \colhead{} & \colhead{} \\
\colhead{} & \colhead{Morphology} & \colhead{Density (D8)} & \colhead{$f_{SF}$} & \colhead{$N_{clust}$} & \colhead{$N_{gal}$} \\
\colhead{} & \colhead{(1)} & \colhead{(2)} & \colhead{(3)} & \colhead{(4)} & \colhead{(5)}}
\startdata
(1) & All & $< 8$ & $0.244 \pm 0.011$ & 68 & 1948 \\
(2) & All & $\ge 8$ & $0.202 \pm 0.014$ & 38 & 2192 \\
\hline
(3) & Filament & $5 \le \textnormal{D}8 < 8$ & $0.258 \pm 0.033$ & 13 & 467 \\
(4) & Filament & $\ge 8$ & $0.166 \pm 0.019$ & 16 & 924 \\
(5) & Spider & $5 \le \textnormal{D}8 < 8$ & $0.231 \pm 0.016$ & 24 & 922 \\
(6) & Spider & $\ge 8$ & $0.229 \pm 0.016$ & 22 & 1268
\enddata
\label{tab:SFfracsTab}
\end{deluxetable}

In this section, we investigate how cluster $f_{SF}$ is related to both the density of the clusters' surrounding environment, and the morphology of the superclusters in which the clusters reside.  For convenience, all $f_{SF}$ values discussed can be found in Table~\ref{tab:SFfracsTab}, with the following columns: (1) morphology of supercluster; (2) environmental density (D8); (3) $f_{SF}$; (4) number of clusters; and (5) number of galaxies.

\begin{figure}[t]
\begin{center}
\includegraphics[scale=0.63]{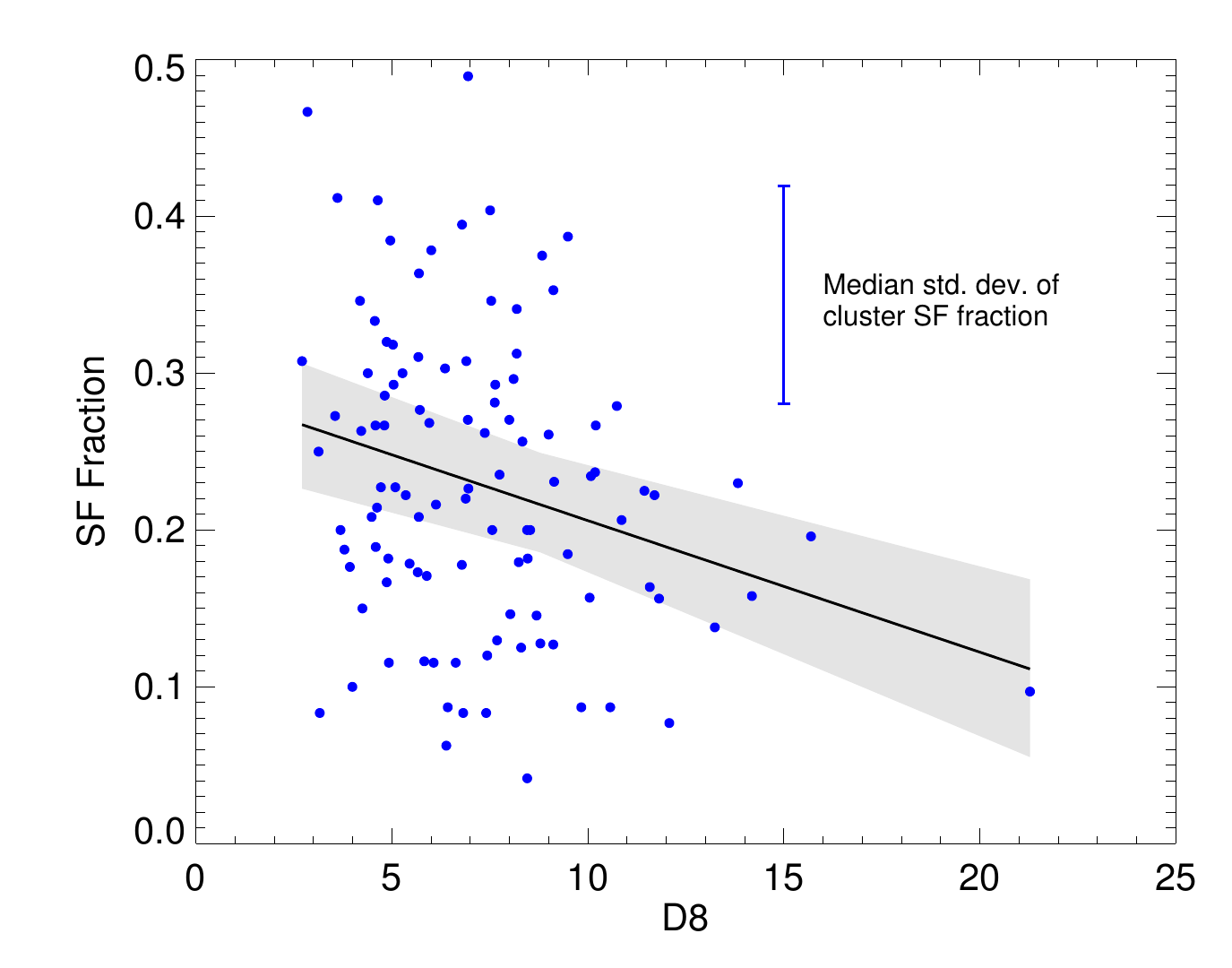}
\caption{$f_{SF}$ versus D8.  Blue points represent individual clusters, and the gray region represents a $1\sigma$ error on the best fit solid line. The error bar is the median standard deviation of each individual cluster's $f_{SF}$. In general, clusters in lower-density environments exhibit higher values of $f_{SF}$.}
\label{fig:SFvsD8Fig}
\end{center}
\end{figure}

In Figure~\ref{fig:SFvsD8Fig}, we plot $f_{SF}$ as a function of D8. Each blue point represents a cluster, and the best fit line is calculated via a linear regression of the cluster values. The gray region represents a $1\sigma$ error on the best fit, which is calculated by performing a bootstrap resampling of all clusters, recalculating the best fit line each time, and taking the standard deviation of the resulting slopes. The error bar represents the median of the standard deviations of each individual cluster's $f_{SF}$. Each cluster's $f_{SF}$ standard deviation is calculated by resampling the galaxies in the cluster, determining a new $f_{SF}$ each time, and taking the standard deviation of these $f_{SF}$ values.

The slope of this relation, $-0.008 \pm 0.002$, is negative at the 99.9\% confidence level with a significance of approximately $3.5\sigma$, indicating that a weak but significant inverse correlation exists between $f_{SF}$ and the density of the supercluster environment. We also calculate the average cluster $f_{SF}$ at lower large-scale densities ($\textnormal{D}8 < 8$; row 1 of Table~\ref{tab:SFfracsTab}) and in high-density supercluster cores ($\textnormal{D}8 \geq 8$; row 2). In low-density areas, we find the $f_{SF}$, $0.244 \pm 0.011$, is higher than that in high-density cores, $0.202 \pm 0.014$, a difference that is significant to 99\% confidence (as determined through bootstrap resampling). These results suggest that, in general, there exist higher values of $f_{SF}$ in clusters in low-density large-scale environments than in high-density cores of superclusters. We note as an aside that, if we remove the highest-density cluster with $\textnormal{D}8 > 20$ from our analysis, the slope of our relation remains negative with a significance at the 99.7\% confidence level.

We test whether differences in cluster mass could be the cause of the observed correlation between SF and D8. Many studies find decreasing SF with increasing cluster mass \citep[e.g.,][]{Finn2005, Homeier2005, Weinmann2005, Poggianti2006, Koyama2010}, while others find no such correlation \citep[e.g.,][]{Goto2005, Popesso2007, Balogh2010, Chung2011}. As a proxy for total halo mass, \citetalias{Cohen2014} used the observed stellar mass of cluster galaxies, $M_{*}$, obtained from the MPA/JHU VAGC \citep{Tremonti2004}. We use a similar metric, but multiply $M_{*}$ by the ratio of $L_{tot}$ to $L_{obs}$ to obtain an estimated total cluster stellar mass, $M_{*}^{tot} = M_{*} \times (L_{tot}/L_{obs})$.

Our method to test the effect of cluster mass is as follows. In short, in each bin of D8 we weight the clusters to have the same $M_{*}^{tot}$ distribution as the sample as a whole, and use these weights to calculate measurement errors for our linear regression. This effectively removes any effect of cluster mass on our $f_{SF}$ measurements. First, we calculate each cluster's $M_{*}^{tot}$ and determine the normalized distribution of these cluster masses. Next, for each bin of D8, we weight the bin's $M_{*}^{tot}$ values so their normalized distribution matches that of our entire sample. Each cluster is assigned the weight of its $M_{*}^{tot}$ bin. Finally, we apply these weights to the galaxies in our linear regression analysis. We find that the slope of our relation actually becomes slightly more negative, decreasing to $-0.11 \pm 0.003$, when controlling for cluster mass. Furthermore, the significance of the correlation increases slightly to $3.7\sigma$. This suggests that a relation between large-scale density and cluster mass is not the cause of the observed correlation between SF and D8. We also perform the same weighting procedure using \emph{r}-band luminosity (both $L_{obs}$ and $L_{tot}$) as a proxy for halo mass, and the results remain the same. We further note that, when plotting $f_{SF}$ as a function of cluster mass, we observe no correlation. Finally, we perform the same weighting procedure using number of cluster galaxies instead of $M_{*}^{tot}$. In this case, the significance of the correlation drops slightly, but still remains above $3\sigma$.

\begin{figure}[t]
\begin{center}
\includegraphics[scale=0.63]{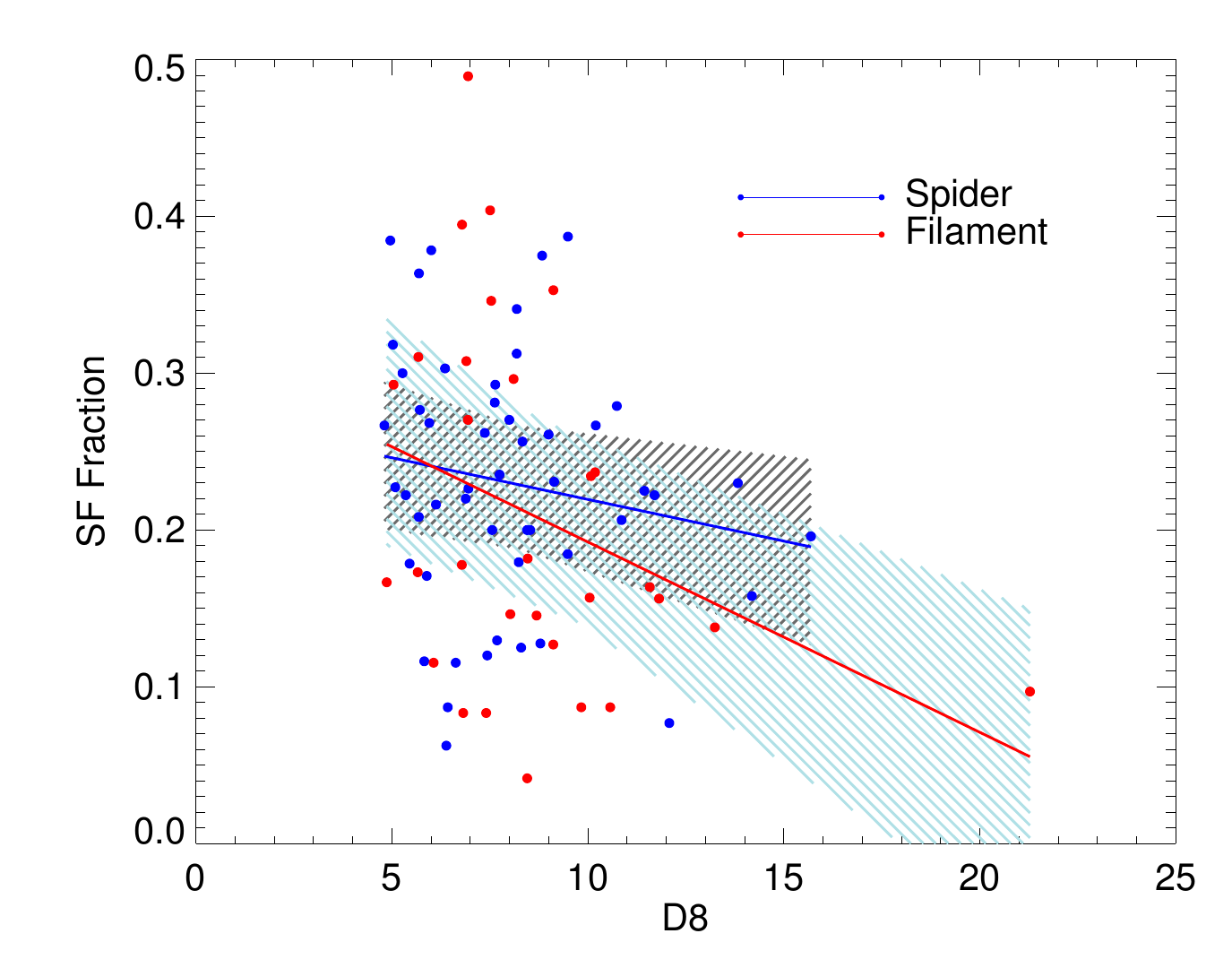}
\caption{$f_{SF}$ versus D8 in spider (blue, right-hatched) and filament (red, left-hatched) superclusters. Points represent individual clusters, and the hatched regions represent $1\sigma$ errors on the best fit lines. The inverse correlation between $f_{SF}$ and D8 is predominantly due to filament superclusters. Also, we observe higher $f_{SF}$ values in spider superclusters than filament superclusters at high environmental densities.}
\label{fig:SFvsD8MorphFig}
\end{center}
\end{figure}

We now test this relationship in superclusters of spider and filament morphology separately. Note that we only include clusters within superclusters, i.e., with $\textnormal{D}8 > 5$. Figure~\ref{fig:SFvsD8MorphFig} shows $f_{SF}$ as a function of D8 for clusters in spider (blue, right-hatched) and filament (red, left-hatched) superclusters. The hatched regions represent $1\sigma$ errors on the best fit lines, determined, as in Figure~\ref{fig:SFvsD8Fig}, by bootstrapping over clusters of each type. Interestingly, we observe the same inverse correlation between $f_{SF}$ and D8 only for filament superclusters: the slope of this relation is negative at the 99.8\% confidence level, and $f_{SF}$ at lower densities, $0.258 \pm 0.033$, is higher than that at higher densities, $0.166 \pm 0.019$, with greater than 99\% confidence (rows 3 and 4 of Table~\ref{tab:SFfracsTab}, respectively). In spider superclusters, there is no significant correlation between $f_{SF}$ and D8.

We also examine $f_{SF}$ in clusters in spider and filament superclusters at high densities (rows 4 and 6 of Table~\ref{tab:SFfracsTab}, respectively).  The value of $f_{SF}$ in spider superclusters with $\textnormal{D}8 > 8$, $0.229 \pm 0.016$, is higher than that in filament superclusters with $\textnormal{D}8 > 8$, $0.166 \pm 0.019$, with greater than 99\% confidence.  This difference is apparent in Figure~\ref{fig:SFvsD8MorphFig}. In low density outskirts (rows 3 and 5 of Table~\ref{tab:SFfracsTab}), there is no difference in $f_{SF}$ between clusters in spider and filament superclusters.

The specifics of the FoF algorithm used by \citet{Tempel2012} introduces a complication into our analysis. As the FoF algorithm builds a given cluster, the higher density of galaxies within superclusters makes it easier for the algorithm to include galaxies at larger cluster radii. Since galaxies on the outskirts of clusters typically exhibit more SF, this could artificially enhance cluster $f_{SF}$ values at higher supercluster densities. We test for this possibility in two ways:

\begin{enumerate}[topsep=0pt,itemsep=-1ex,partopsep=1ex,parsep=1ex]
\item We first measure $f_{SF}$ against the ratio of virial radius ($r_{vir}$, derived as the projected harmonic mean radius by \citealt{Tempel2012}) to $L_{obs}$, serving as a proxy for a measurement of cluster radius based solely on cluster mass \citep[e.g.,][]{Yang2007}. Clusters with high $r_{vir}$ for their mass-derived radii (via $L_{obs}$) could have enhanced values of $f_{SF}$ due to galaxies in outskirts included by the FoF algorithm. We find no correlation between these quantities.
\item Second, we measure $f_{SF}$ against the surface density of cluster galaxies ($L_{obs}/\pi r_{vir}^2$). Clusters with lower surface densities may be artificially expanded by the FoF algorithm, including galaxies in outskirts with higher SF and thus exhibiting enhanced values of $f_{SF}$. Again, we find no correlation between these quantities.
\end{enumerate}

We perform these tests not only with $L_{obs}$ as a proxy for cluster radius and mass, but also with $L_{tot}$ and $M_{*}^{tot}$. The results of the tests suggest that any extended tails of galaxies included in clusters due to the FoF algorithm are not artificially enhancing cluster $f_{SF}$.

All of these results suggest that 1) there is a significant inverse correlation between $f_{SF}$ and D8, dominated by clusters in filament superclusters; and 2) in high-density cores of superclusters, spider superclusters exhibit higher values of $f_{SF}$ than filament superclusters.

\begin{deluxetable}{lcccc}
\tablecolumns{4}
\tablewidth{0pc}
\tablehead{
\colhead{Variable} & \colhead{PC1} & \colhead{PC2} & \colhead{PC3} & \colhead{PC4}}
\startdata
$f_{SF}$ & $\hphantom{-}0.173$ & $\hphantom{-}0.730$ & $-0.662$ & $-0.008$ \\
$-\log(p_{\Delta})$ & $-0.302$ & $\hphantom{-}0.664$ & $\hphantom{-}0.656$ & $-0.195$ \\
D8 & $-0.650$ & $-0.159$ & $-0.336$ & $-0.662$ \\
$L_{obs} [10^{10} h^{-2} L_{\odot}]$ & $-0.675$ & $\hphantom{-}0.042$ & $-0.139$ & $\hphantom{-}0.723$ \\
\hline
Std. dev. & $\hphantom{-}1.403$ & $\hphantom{-}1.161$ & $\hphantom{-}0.714$ & $\hphantom{-}0.415$ \\
Prop. of var. & $\hphantom{-}0.492$ & $\hphantom{-}0.337$ & $\hphantom{-}0.127$ & $\hphantom{-}0.043$ \\
Cum. prop. & $\hphantom{-}0.492$ & $\hphantom{-}0.830$ & $\hphantom{-}0.957$ & $\hphantom{-}1.000$
\enddata
\label{tab:PCATab}
\end{deluxetable}

\subsection{Principal component analysis}
\label{sec:PCASec}

As discussed in \S\ref{sec:PCADataSec}, we use PCA to determine any correlations between cluster $f_{SF}$, amount of cluster substructure, density of supercluster environment, and total cluster mass. We use two measurements of amount of substructure from \citetalias{Cohen2014}: number of components; and the results from the $\Delta$ test, which in this case is the negative of $\log(p_{\Delta})$. We also use two proxies for total cluster mass: \emph{r}-band luminosity (both $L_{obs}$ and $L_{tot}$) and $M_{*}^{tot}$.

Our PCA results are consistent whether we use $L_{obs}$, $L_{tot}$, or $M_{*}^{tot}$ as a proxy for total cluster mass. Furthermore, our results remain the same whether we use $-\log(p_{\Delta})$ or number of components as a measure of amount of substructure. Thus, we present and discuss only the results when using $L_{obs}$ and $-\log(p_{\Delta})$. Table~\ref{tab:PCATab} displays the results of our analysis. It shows the values of the four PCs for our four variables; and the standard deviation, proportion of variance, and cumulative proportion for these PCs.

The cumulative proportion shows that the first two PCs account for 83\% of the variance in these cluster properties, with each PC being equally important. Thus, we will focus primarily on the first two PCs. The PC1 values of D8 and $L_{obs}$ are close in magnitude and of the same sign, confirming the correlation between these two variables. Since the value of $f_{SF}$ is of opposite sign, this suggests that $f_{SF}$ is weakly anti-correlated with D8 and $L_{obs}$. Furthermore, the PC2 value of D8 -- also of opposite sign to $f_{SF}$ -- is approximately four times larger than that of $L_{obs}$. This suggests that D8 rather than $L_{obs}$ is more strongly related to $f_{SF}$. This agrees with our analysis in \S\ref{sec:PairwiseSec}, where we confirm that differences in $M_{*}^{tot}$ (i.e., cluster \emph{r}-band luminosity) are not the cause of the observed correlation between $f_{SF}$ and D8.

The PC2 values of $f_{SF}$ and amount of substructure are of similar magnitude and of the same sign, confirming the direct correlation between these variables found in \citetalias{Cohen2014}. While the PC1 values of these variables are of opposite sign, the correlation suggested by the PC2 values is more robust: the values of these variables along PC2 are closer in magnitude to each other than those along PC1; and the D8 and $L_{obs}$ values along PC2 are much lower than those along PC1. These values, compared to the others discussed, suggest that $f_{SF}$ is probably most strongly related to amount of substructure than the other variables discussed.

Finally, the PC1 values of D8 and amount of substructure suggest a direct correlation between these variables, which agrees with the findings in \citetalias{Einasto2012a} (though they admit that this correlation is weak). Furthermore, PC1 also shows that amount of substructure is also correlated with $L_{obs}$. Intuitively, this is expected: a richer cluster will have a higher luminosity and more opportunity for substructure to be detectable. This effect does act counter to the main result of this paper -- the inverse correlation between D8 and $f_{SF}$ -- and the result of \citetalias{Cohen2014} -- the direct correlation between substructure and $f_{SF}$ (see Figures 5 and 6 in that paper). In other words, as D8 and luminosity increase, the results from this paper and \citetalias{Cohen2014} suggest that $f_{SF}$ (and thus amount of substructure) should decrease, not increase. This, however, bolsters the correlation we find between D8 and $f_{SF}$ -- it must be significant enough to counter the weak correlation between D8 and amount of substructure.

\section{Discussion}
\label{sec:DiscussionSec}
We find a significant inverse correlation between the density of supercluster environment and the amount of SF within galaxy clusters. While this could in principle be an indirect result of a correlation between supercluster density and cluster substructure, we find this not to be the case. Rather, both cluster substructure and supercluster environment are independently related to a cluster's SF, and, while these effects oppose each other, the influence of cluster substructure appears stronger than that of supercluster environment. These results are not simply due to the correlation between D8 and cluster mass, luminosity, or richness.

We also find that supercluster morphology is important in affecting cluster SF: the relation between supercluster density and SF is observed only in filament rather than spider superclusters. Furthermore, SF in spider superclusters is higher at high densities compared to filament superclusters. When we consider these differences between filament and spider superclusters, from the complexity of effects explained above emerges a coherent picture. Spider superclusters have richer inner structure, and are dynamically younger, than filament superclusters \citepalias[e.g.,][]{Einasto2012b}. We expect to find more SF in dynamically younger systems, and we indeed see this in the high-density cores of spider superclusters.

In galaxy clusters, more structure indicates a less relaxed, younger system (e.g., \citealt{Bird1993}; \citealt{Knebe2000}; \citetalias{Cohen2014}; \citealt{Cohen2015}). Such clusters are more likely to live in superclusters with richer inner structure where group mergers occur more easily than in superclusters with simple inner structure. Thus, combining our results from this work with those of \citetalias{Cohen2014}, we can explain in more detail the effects of cluster substructure and supercluster environment on SF. As clusters form hierarchically from smaller groups, the dynamically younger systems exhibit more SF, since the SF in these systems has had less time to be quenched by various gravitational and hydrodynamical processes (see \citealt{Boselli2006} for a review of such mechanisms). Thus, it is more likely to find high SF in clusters that find themselves in the high-density environments of spider superclusters than filament superclusters. Additionally, this shows that high-density cores of superclusters are a special environment for clusters. For instance, they may be collapsing \citep{Einasto2015, Gramann2015, Einasto2016}, possibly affecting the properties of galaxy clusters and their galaxy populations. This interesting result of our study emphasizes the role of supercluster morphology in shaping the properties of galaxies and groups/clusters in them.

The main result of this work agrees with \citet{Lietzen2012}, who found that more elliptical galaxies are found in the higher-density environments of superclusters than at lower densities. Furthermore, \citet{Luparello2013} used the galaxy spectra parameter $D_{n}4000$ to show that galaxies in groups in superclusters are systematically \emph{older} than those in lower-density environments. They found that this result holds even though the groups themselves have higher velocity dispersions and are therefore dynamically \emph{younger} than groups elsewhere. These results agree well with the interpretation from this work explained above.

We note that \citet{Costa2013} found no correlation between the mean stellar ages of superclusters and the shape parameter of superclusters. This is not in conflict with our results, since we used information about the inner structure of superclusters to divide them into two morphological classes, while \citet{Costa2013} only used the shape parameter to characterize the outer shape of superclusters. \citet{Einasto2014} showed, in agreement with \citet{Costa2013}, that galaxy content of superclusters depends only weakly on the overall shape of superclusters.

Our results, while significant, still exhibit substantial scatter. This owes to the complicated dynamics affecting cluster $f_{SF}$, many aspects of which are discussed here. One variable we have not taken into account is the stage of formation of a cluster or supercluster. Studies have shown that clusters with similar degrees of apparent substructure can exhibit different $f_{SF}$ measurements due to different stages of cluster merger activity \citep[e.g.,][]{Hwang2009}. Future studies including the affects of cluster or supercluster age could succeed in reducing the scatter in the results discussed in this work.

\acknowledgements

We thank the SDSS team, as well as the MPA/JHU and NYU researchers, for the publicly available data releases and VAGCs.  Funding for the SDSS and SDSS-II has been provided by the Alfred P. Sloan Foundation, the Participating Institutions, the National Science Foundation, the U.S. Department of Energy, the National Aeronautics and Space Administration, the Japanese Monbukagakusho, the Max Planck Society, and the Higher Education Funding Council for England. The SDSS Web Site is http://www.sdss.org/.  The SDSS is managed by the Astrophysical Research Consortium for the Participating Institutions. The Participating Institutions are the American Museum of Natural History, Astrophysical Institute Potsdam, University of Basel, University of Cambridge, Case Western Reserve University, University of Chicago, Drexel University, Fermilab, the Institute for Advanced Study, the Japan Participation Group, Johns Hopkins University, the Joint Institute for Nuclear Astrophysics, the Kavli Institute for Particle Astrophysics and Cosmology, the Korean Scientist Group, the Chinese Academy of Sciences (LAMOST), Los Alamos National Laboratory, the Max-Planck-Institute for Astronomy (MPIA), the Max-Planck-Institute for Astrophysics (MPA), New Mexico State University, Ohio State University, University of Pittsburgh, University of Portsmouth, Princeton University, the United States Naval Observatory, and the University of Washington. ME and JV were supported by the ETAG project
IUT26-2 and by the Centre of Excellence ``Dark side of the Universe" (TK133) financed by the European Union through the European Regional Development Fund.

\end{document}